%

\input iopppt.tex



\catcode`@=11
\newwrite\auxfile
\newwrite\xreffile
\newif\ifxrefwarning \xrefwarningtrue
\newif\ifauxfile
\newif\ifxreffile
\def\testforxref{\begingroup
    \immediate\openin\xreffile = \jobname.aux\space
    \ifeof\xreffile\global\xreffilefalse
    \else\global\xreffiletrue\fi
    \immediate\closein\xreffile
    \endgroup}
\def\testforaux{\begingroup
    \immediate\openin\auxfile = \jobname.aux\space
    \ifeof\auxfile\global\auxfilefalse
    \else\global\auxfiletrue\fi
    \immediate\closein\auxfile
    \endgroup}
\def\openreffile{\immediate\openout\auxfile = \jobname.aux}%
\def\readreffile{%
    \testforxref
    \testforaux
    \ifauxfile
       \begingroup
         \@setletters
         \input \jobname.aux
       \endgroup
    \else
\message{No cross-reference file existed, some labels may be undefined}%
    \fi\openreffile}%
\def\@setletters{%
    \catcode`_=11 \catcode`+=11
    \catcode`-=11 \catcode`@=11
    \catcode`0=11 \catcode`1=11
    \catcode`2=11 \catcode`3=11
    \catcode`4=11 \catcode`5=11
    \catcode`6=11 \catcode`7=11
    \catcode`8=11 \catcode`9=11
    \catcode`(=11 \catcode`)=11
    \catcode`:=11 \catcode`'=11
    \catcode`&=11 \catcode`;=11
    \catcode`.=11}%
\gdef\el@b{\eqlabel}
\gdef\sl@b{\seclabel}
\gdef\tl@b{\tablabel}
\gdef\fl@b{\figlabel}
\def\l@belno{\ifx\labeltype\el@b 
   \let\labelno=\en\def\@label{\eqlabel}%
   \else\let\labelno=\ignorespaces
   \ifx\labeltype\sl@b \def\@label{\seclabel}%
   \else\ifx\labeltype\tl@b \def\@label{\tablabel}%
   \else\ifx\labeltype\fl@b \def\@label{\figlabel}%
   \else\def\@label{\seclabel}\fi\fi\fi\fi}
\def\label#1{\l@belno\expandafter\xdef\csname #1@\endcsname{\@label}%
    \immediate\write\auxfile{\string
    \gdef\expandafter\string\csname @#1\endcsname{\@label}}%
    \labelno}%
\def\ref#1{%
    \expandafter \ifx \csname @#1\endcsname\relax
    \message{Undefined label `#1'.}%
    \expandafter\xdef\csname @#1\endcsname{(??)}\fi 
    \csname @#1\endcsname}%
\readreffile


\def\bibitem#1{\global\advance\refno by 1%
    \immediate\write\auxfile{\string
    \gdef\expandafter\string\csname #1@\endcsname{\the\refno}}%
    \rf{[\the\refno]}}%
\def\bitem[#1]#2{\immediate\write\auxfile{\string
    \gdef\expandafter\string\csname #2@\endcsname{#1}}%
    \rf{[#1]}}%
\def\cite#1{\hbox{[\splitarg{#1}]}}%
\def\splitarg#1{\@pt#1,\@ptend}%
\def\@pt#1,#2\@ptend{\ifempty{#1}\else
    \@pttwo #1\@pttwoend
    \ifempty{#2}\else\sp@cer\@pt#2\@ptend\fi\fi}%
\def\@pttwo#1\@pttwoend{\expandafter
    \ifx \csname#1@\endcsname\@pttwoend\else
    \@ifundefined{#1}{{\bf ?}%
    \message{Undefined citation `#1' on page 
    \the\pageno}}{\csname#1@\endcsname}\fi}%
\def\@pttwoend{@@@@@}%
\def\sp@cer{,\nobreak\thinspace}%
\def\ifempty#1{\@ifempty #1\@xx\@xxx}%
\def\@ifempty#1#2\@xxx{\ifx #1\@xx}%
\def\@xx{@@@@}%
\def\@xxx{@@@@}%
\long\def\@ifundefined#1#2#3{\expandafter\ifx\csname
  #1@\endcsname\relax#2\else#3\fi}%
\catcode`@=12
%


\def\CMP{{\it Comm. Math. Phys.}}
\def\ei{{\rm i}}

\def\Ph{\nthorn}
\def\D{\nedth}
\parindent=0pt

\def\pmb#1{\setbox0=\hbox{#1}%
  \kern-.025em\copy0\kern-\wd0
  \kern.05em\copy0\kern-\wd0
  \kern-0.025em\raise.0433em\box0 }
\def\half{{\scriptstyle {1 \over 2}}}
\def\third{{\scriptstyle {1 \over 3}}}
\def\quarter{{\scriptstyle {1 \over 4}}}
\def\o{{\pmb o}} 
\def\i{{\pmb{$\iota$}}}
\let\gt=\mapsto
\let\la=\lambda

\def\ob{{\bar\o}}
\def\ib{{\bar\i}}
\def\T{{\bf T}}
\def\bp{{\bf p}}
\def\bq{{\bf q}}
\def\0{\pmb 0} 
\def\1{\pmb 1} 
\def\2{\pmb 2} 
\def\3{\pmb 3} 
\def\bPhi{{\pmb{$\Phi$}}}
\def\>{\phantom{A}}
\def\sym{{\sum_{sym}}}
\def\thorn{\hbox{\rm I}\kern-0.32em\raise0.35ex\hbox{\it o}}
\def\edth{\hbox{$\partial$\kern-0.25em\raise0.6ex\hbox{\rm\char'40}}}

\def\nedth{{\pmb\edth}}
\def\nedthp{{\pmb\edth'}}
\def\thornp{\thorn'}
\def\nthorn{{\pmb\thorn}}
\def\nthornp{{\pmb\thorn'}}
\def\edthp{\edth'}

\def\cd{{\cal D}}
\def\boeta{{\pmb{$\eta$}}}
\def\blambda{{\pmb{$\lambda$}}}
\def\etad{\boeta_{C_1\dots C_N C'_1\dots C'_{N'}}}
\def\p{{\bf p}}
\def\q{{\bf q}}
\def\K{{\bf K}}
\def\R{{\bf R}}
\def\S{{\bf S}}
\def\T{{\bf T}}
\def\I{{\bf I}}
\def\la{\lambda}
\def\lab{\bar\lambda}



\pptstyle

\jl{6}

\title{Integration using invariant operators: Conformally flat
  radiation metrics}[Integration using invariant operators]
\author{S B Edgar\dag\   and J A Vickers\ddag}[S B Edgar and J A Vickers]
\address{\dag Department of Mathematics, Link\"oping University,
Link\"oping, Sweden S-581~83}
\address{\ddag Department of Mathematics, University of Southampton,
Southampton, SO17~1BJ, UK}
\beginabstract
A new method is presented for obtaining the general conformally flat
radiation metric by using the differential operators of Machado Ramos
and Vickers (a generalisation of  those of Geroch, Held and Penrose) which 
are invariant under null rotations and rescalings. The solution is
found by constructing involutive tables of these derivatives applied
to the quantities which arise in the Karlhede classification of this
class of metrics.
\endabstract
\submitted
\date
\pacs{0420, 1127}


\section{Introduction}
In a recent paper [1] Machado Ramos and Vickers introduced some new
operators which are invariant under null rotations. In a subsequent
paper [2] this was generalised to incorporate spin and boost
transformations so that the resulting formalism depends only upon a
choice of a single null direction. Not surprisingly this formalism
combines many features of the GHP [3] and null rotation invariant
formalisms. In this new formalism the role of the spin coefficients
$\kappa$, $\sigma$, $\rho$ and $\tau$ is taken up by spinor quantities
$\K$, $\S$, $\R$ and $\T$ given by
$$\eqalign{
\K&=\kappa \cr
\S_{A^\prime}&=\sigma \ob_{A^\prime}-
\kappa\ib_{A^\prime} \cr
\R_A&=\rho\o_A-\kappa\i_A \cr
\T_{AA^\prime}&= \tau \o_A\ob_{A^\prime}- \rho
\o_A\ib_{A^\prime}- \sigma \i_A\ob_{A^\prime}+
\kappa\i_A\ib_{A^\prime} \cr
}
$$
Under a transformation of the spin frame given by
$$
\o^A\gt\la\o^A \qquad \i^A\gt\la^{-1}\i^A+\bar a \o^A 
$$
these transform as
$$\eqalign{
\K&\gt \la^3\lab \K \cr
\S_{A'}&\gt \la^3\S_{A'} \cr
\R_A&\gt \la^2\lab \R_A \cr
\T_{AA'}&\gt \la^2 \T_{AA'} \cr
}
$$
They are therefore invariant under null rotations and have weight $\{\bp, \bq\}$
under spin and boost transformations given by
$$\eqalign{
\K&:\quad\{\3, \1\} \cr
\S&:\quad\{\3, \0\} \cr
\R&:\quad\{\2, \1\} \cr
\T&:\quad\{\2, \0\} \cr
}
$$
The role of the differential operators $\thorn$, $\edth$, $\thornp$
and $\edthp$ is played by new differential operators $\nthorn$,
$\nedth$, $\nthornp$ and $\nedthp$ which act on properly weighted
symmetric spinors to produce symmetric spinors of different valence
and weight. These operators may all be defined in terms of an
auxiliary differential operator $\cd_{ABA'B'}$ which is defined by
$$\eqalign{
\cd_{ABA'B'}\etad &=\o_A\ob_{A'}\nabla_{BB'}\etad \cr
&-(\p\ob_{A'}\nabla_{BB'}\o_A+\q\o_A\nabla_{BB'}\ob_{A'})\etad \cr
} \label{di}
$$
where $\boeta$ has weight $\{\p,\q\}$.
\medskip\noindent
The original motivation for introducing such operators was to improve
on the derivative bounds for the Karlhede classification. However it
was also hoped that such a formalism would prove useful in finding
solutions to Einstein's equations which are invariant under null
rotations. The use of the GHP formalism to find exact solutions was
pioneered by Held [4,5,6] and in the past few years has been applied
by a number of authors including Edgar and Ludwig
[7,8,9]. In particular in [9] they demonstrated how the GHP formalism
could be used to obtain the complete class of conformally flat
radiation metrics. Their method consisted of manipulating the complete system
of GHP equations until they were reduced to a complete involutive set of tables
for the action of the four GHP operators on four  functionally
independent $\{0,0\}$ weighted real scalars, and on one  non-trivially
weighted  complex scalar. As Held had emphasised once such a
`complete set of tables' was obtained the problem was solved; by considering as
coordinates  the four real
$\{0,0\}$ weighted scalars    it was straightforward to write down, directly
from their respective tables, the tetrad and hence the metric  explicitly.
\medskip\noindent
Another approach to the construction of exact solutions which was
originally suggested by Karlhede and Lindstr\"om [10], is to apply the
techniques used in classifying equivalent metrics in reverse and
construct a geometry from a set of elements representing the Riemann
tensor and some of its covariant derivatives. This is indeed possible
provided certain integrability conditions are satisfied [11,
12]. Furthermore in a number of papers Bradley and Marklund [13, 14]
have actually used the method to construct a class of locally
rotationally symmetric perfect fluid spacetimes. In the present paper
we will combine the ideas used in these two approaches by performing
the integration using a systematic application of the commutators of
the invariant differential operators to the functional information
obtained at each order of the Karlhede classification.
\medskip\noindent
Since the formalism of Machado Ramos and Vickers is invariant under
null rotations it should, in principal, be ideally suited to
describing the conformally flat radiation metrics. However since the
formalism involves symmetric spinors rather than scalars, it was not
completely clear how one would carry out the integration in
practice. In this paper we demonstrate how the formalism may be used to
find all the metrics in the class. The key to the method is to extract
$\{0,0\}$ weighted scalars from spinor quantities. This can happen in
two possible ways. 
\medskip\noindent
Firstly if one has a spinor field $\boeta_A$ of weight $\{\1, \0\}$
such that $\boeta_A\o^A=0$ then $\boeta_A$ has the form
$$
\boeta_A=\eta\o_A
$$
for some scalar field $\eta$. Since $\o_A$ has weight $\{\1,\0\}$ then
$\eta$ must be a weight $\{0,0\}$ scalar. This procedure works for a
general type $(N,N')$-spinor field $\boeta$ of weight $\{{\bf N},{\bf N}'\}$ with the
property that $\boeta\cdot\o=0$ since one must then have
$$
\boeta_{A_1\ldots
A'_{N'}}=\eta\o_{A_1}\ldots\o_{A_N}\ob_{A'_1}\ldots\ob_{A'_{N'}}
$$
where $\eta$ is a weight $\{0,0\}$ scalar.
\medskip\noindent
The second way in which one can extract scalar fields, is to identify in
some invariant way a weight $\{-\1,\0\}$ spinor field $\blambda_A$ which
is not proportional to $\o_A$, i.e for which $\blambda_A \o^A=-\lambda
\neq 0$. In this case one may {\it define} $\i_A$ by
$$
\blambda_A=\lambda\i_A
$$
Furthermore one may then contract any other spinor field with $\i^A$
to construct {\it invariant} scalar fields.
\medskip\noindent
For a conformally flat radiation solution one may choose $\o_A$ to be
aligned with the propagation direction of the radiation and we shall
see that this means that the
Ricci spinor $\bPhi_{ABA'B'}$ takes the form
$$
\bPhi_{ABA'B'}=\Phi\o_A\o_B\ob_{A'}\ob_{B'}
$$
where $\Phi$ is a (real) scalar field of weight $\{-2,-2\}$. Other
scalar fields may be obtained by taking the covariant derivatives 
of $\bPhi_{ABA'B'}$. It turns out that the components of $\bPhi_{AA'BB';CC'}$
may be given in terms of $\thorn \Phi$, $\edth \Phi$, $\edth'\Phi$ and
$\thorn'\Phi$. Because of the contracted Bianchi identities one finds
that $\thorn \Phi=0$ but that $\edth \Phi$ is a type $\{-1,-3\}$
scalar which despite appearances {\it does not} depend upon the choice of
$\i_A$ (in fact the Bianchi identities show that
$\edth\Phi=\tau\Phi$). 
$\thorn' \Phi $ on the other hand {\it does} depend upon the choice
of $\i_A$ and generically one can choose $\i_A$ to make $\thorn' \Phi$
vanish. Thus for such metrics the curvature and its first derivative
provide two scalar fields; $\Phi$  and $\edth\Phi$ which are invariant
under null rotations. However these quantities have non-trivial weights
and are not invariant under spin and boost transformations. We
therefore construct an algebraic combination of $\Phi$ and $\edth\Phi$,
which we denote $A$, which is a real scalar field of $\{0,0\}$ weight
and hence invariant under both null rotations and spin and boosts.
The remaining two pieces of information in $\Phi$ and $\edth\Phi$ may be encoded in
a complex field $P$, of
modulus one half, which has spin weight one and boost weight zero and a real
field $Q$ which has boost weight one and spin weight zero. For convenience we may
combine these into a single complex field $\bar P Q$. 
\medskip\noindent
In the usual Karlhede approach to the classification one starts by
choosing the spin frame so that $\Phi=1$. Since $\Phi$ is
real this amounts to fixing the boost freedom in $\o_A$. At the next
stage  the new functional information is provided by the complex
scalar field $\tau$. The rotational freedom of $\o_A$ may now be
fixed, for instance by demanding the $\tau$ is real, and the real valued
scalar field that one obtains provides the first piece of functional
information. The only other information one obtains from the first
derivative is contained in $\thorn'\Phi$, but as we have seen above
this may be set to zero by partially fixing the null rotation freedom. Thus
generically at first order one has
one piece of functional information. However in the present approach
we wish to work with operators which are invariant under both spin and
boosts and null rotations. Thus rather than fixing the gauge by
demanding canonical forms for $\Phi$ and its covariant derivative 
we will instead use
the information contained in them to construct $A$, $P$ and $Q$. We
regard $P$ and $Q$ as
gauge fields, as the final answer does not depend upon them, while the
invariant scalar field $A$ contains the functional information. In the
same way we will not fix the null rotation freedom by putting the
first derivative into a canonical form, but instead,
introduce in a natural way a spinor field $\I_A$  of weight $\{-1,0\}$
which satisfies $\o_A\I^A=1$. Again this spinor field contains only
gauge information and the final answer does not depend upon
it. This should be contrasted with the usual NP integration procedure
where one uses the spin coefficients and their derivatives to
completely fix the frame, or the GHP integration procedure where one
fixes the null rotation freedom.
\medskip\noindent 
The procedure then is to manipulate the complete set of spinor equations in
the formalism of Machado Ramos and Vickers   in an analogous manner to that
followed in the GHP formalism in [9]. In general  this will involve
weighted spinor fields, but by using
$\bar PQ$, and $\I_A$ (and their conjugates) one can 
extract  {\it $\{0,0\}$ weighted scalar
fields}. Eventually one obtains a `complete set of tables', and from the tables
for the four functionally independent $\{0,0\}$ weighted  real scalar fields
the tetrad, and hence the metric, may be obtained in exactly the same way as in
[9].  Although no new exact solutions are produced, this paper demonstrates in
Sections 3 and 4 how to carry out an integration procedure using the new
invariant operators adapted to the symmetry of the
problem. Furthermore our method involves constructing the geometry
solely from the Ricci tensor and its derivatives so that the 
integration procedure essentially carries
out the techniques used in classifying equivalent metrics in reverse. 
In particular it is
clear that the functionally independent scalars, which take on the role of
coordinates in our integration procedure, are simply the invariant scalars
required by the Karlhede classification. This relationship is discussed in
Section 5.

\section{The differential operators and the commutators.}

We begin with an examination of some of the properties of the
differential operators. In particular we need to know the result of
contracting $\nthornp\boeta$ with $\o$ and $\ob$. We start by
rewriting equation \ref{di} in the form
$$\eqalign{
&\!\!\cd_{ABA'B'}\etad =(\thornp\etad)\o_A\o_B\ob_{A'}\ob_{B'} \cr
&\!\!\!\!\!\!\!\!-(\edthp\etad)\o_A\o_B\ob_{A'}\ib_{B'}-(\edth\etad)\o_A\i_B\ob_{A'}
\ob_{B'}\cr
&\!\!\!\!\!\!\!\!-(\thorn\etad)\o_A\i_B\ob_{A'}\ib_{B'}+(\bp\i_A\ob_{A'}\T_{BB'}
+\bq\o_A\ib_{B'}\bar{\T}_{B'B})\etad
}
$$
where $\thornp$, $\edthp$, $\edth$ and $\thorn$ are the ordinary GHP
operators applied to spinors. The new operators are obtained by
contraction with $\o$ and $\ob$, and symmetrizing.
$$\eqalign{
(\nthorn\boeta)_{AC_1\dots C_NA'C'_1\dots C'_{N'}}
&= \sym\o^B\ob^{B'}\cd_{ABA'B'}\etad  \cr
(\nedth\boeta)_{AC_1\dots C_NA'B'C'_1\dots C'_{N'}}
&= \sym\o^B\cd_{ABA'B'}\etad  \cr
(\nedthp\boeta)_{ABC_1\dots C_NA'C'_1\dots C'_{N'}}
&= \sym\ob^{B'}\cd_{ABA'B'}\etad  \cr
(\nthornp\boeta)_{ABC_1\dots C_NA'B'C'_1\dots C'_{N'}}
&= \sym\cd_{ABA'B'}\etad \cr
}
$$
where $\displaystyle\sym$ indicates symmetrization over all free primed and
unprimed indices.
\medskip\noindent
In the case of a scalar field this gives
$$\eqalign{
(\nthornp\eta)_{ABA'B'}&=
(\thornp\eta)\o_A\o_B\ob_{A'}\ob_{B'}
-(\edthp\eta-q\bar\tau\eta)\o_A\o_B\ob_{(A'}\ib_{B')} \cr
&-(\edth\eta-p\tau\eta)\o_{(A}\i_{B)}\ob_{A'}\ob_{B'}
+(\thorn\eta-p\rho\eta-q\bar\rho\eta)\o_{(A}\i_{B)}\ob_{(A'}\ib_{B')}
\cr
&+(p\kappa\i_A\i_B\ob_{(A'}\ib_{B')}+q\bar\kappa\o_{(A}\i_{B)}\ib_{A'}\ib_{B'}
\cr
&-p\sigma\i_A\i_B\ob_{A'}\ob_{B'}-q\bar\sigma\o_A\o_B\i_{A'}\i_{B'})\eta\cr
}\label{thornp}
$$

$$\eqalign{
(\nedthp\eta)_{ABA'}&=
(\edthp\eta)\o_A\o_B\ob_{A'}
-(\thorn\eta-p\rho\eta)\o_{(A}\i_{B)}\ob_{A'}
+(q\bar\sigma\o_A\o_B\i_{A'} \cr
&-p\kappa\i_A\i_B\ob_{A'}-q\bar\kappa\o_{(A}\i_{B)}\ib_{A'})\eta\cr
}\label{edthp}
$$

$$\eqalign{
(\nedth\eta)_{AA'B'}&=
(\edth\eta)\o_A\ob_{A'}\ob_{B'}
-(\thorn\eta-q\bar\rho\eta)\o_{A}\ob_{(A'}\ib_{B')}
+(p\sigma\i_A\ob_{A'}\ob_{B'} \cr
&-p\kappa\i_A\ob_{(A'}\ib_{B')}-q\bar\kappa\o_{A}\ib_{A'}\ib_{B'})\eta\cr
}\label{edth}
$$

$$\eqalign{
(\nthorn\eta)_{AA'}&=
(\thorn\eta)\o_A\ob_{A'}
+(p\kappa\i_A\ob_{A'}-q\bar\kappa\o_{A}\ib_{A'})\eta\cr
}\label{thorn}
$$
Contracting \ref{thornp} with $\ob^{B'}$ gives
$$\eqalign{
(\nthornp\eta)_{ABA'B'}\ob^{B'}&=
\half\{(\nedthp\eta)_{ABA'}-q(\bar\tau\o_A\o_B\ob_{A'}
-\bar\rho\o_{(A}\i_{B)}\o_{A'} \cr
&-\bar\sigma\o_A\o_B\ib_{A'}
+\bar\kappa\o_{(A}\i_{B)}\i_{A'})\eta\} \cr
&=\half\{(\nedthp\eta)_{ABA'}-q\bar{\T}_{A'(A}\o_{B)}\eta\}\cr
}
$$
or in the compacted notation
$$
(\nthornp\eta)\cdot\ob=\half\{(\nedthp\eta)-q\bar\T\eta\} \label{thornp.ob}
$$
Similar calculations give
$$
(\nthornp\eta)\cdot\o=\half\{(\nedth\eta)-p\T\eta\} \label{thornp.o}
$$
$$
(\nedthp\eta)\cdot\o=\half\{(\nthorn\eta)-p\R\eta\} \label{edthp.o}
$$
$$
(\nedth\eta)\cdot\ob=\half\{(\nthorn\eta)-q\bar\R\eta\} \label{edth.ob}
$$
and
$$
(\nthornp\eta)\cdot\o\cdot\ob=\quarter\{(\nthorn\eta)-p\R\eta-q\bar\R\eta\}
\label{thornp.oob}
$$
If $\boeta$ is a spinor the above contractions become more
complicated. For example for a type (1,0)-spinor $\boeta_A$ of weight
$\{\p,\q\}$ we get
$$
(\nthornp\boeta)\cdot\o=\third\{\nthornp(\boeta\cdot\o)+
(\nedthp\boeta)-(\p-1)\T\boeta\}
$$
\medskip\noindent
Although the definition of the differential operators is quite
complicated, the fact that they take symmetric spinors to symmetric
spinors means that one can write down the equations in an index free
notation.

The Ricci equations, Bianchi equations and the commutators for
the general case are given in [2]. This complete system of equations
is completely equivalent to Einstein's equations, and to find solutions to
Einstein's equations this  system will therefore have to be completely
integrated.  This complete system also contains exactly the same information as
the analogous complete systems in the  NP and GHP formalisms respectively.
However, in view of the more complicated nature of the operators in this
formalism, some of the information which resided in the Ricci equations in NP
and/or GHP formalisms is contained within the commutators in this formalism; in
particular these commutators  contain inhomogeneous terms explicitly
dependent on the weight and valence of spinor on which they act.   To
extract all
the information in the commutators we need to apply them to, [15]

(i)\ \ four functionally
independent $\{0,0\}$ weighted real scalars,

(ii)\  one $\{p,q\}$ weighted complex scalar where $p\ne \pm q$,

(iii) one   valence $(1,0)$ spinor, $\I_A$ of  weight $\{{\bf -1},{\bf 0}\}$.

Of course, we can extract all the information by applying the commutators to
different (but essentially equivalent)  combinations of these scalars and
spinor; however the particular choices above are best suited to our integration
procedure since the four $\{0,0\}$ weighted real scalars will become the
coordinates, the complex scalar is given by the gauge field $\bar PQ$,
while the spinor $\I_A$ will be identified with the second dyad
spinor $\i_A$.

For the special case of  the conformally flat
solutions we are considering here there is considerable simplification,
particularly in the Ricci and Bianchi equations. Choosing
$\o_A$ to be aligned with the propagation direction of the radiation means
that the Ricci spinor takes the form
$$
\bPhi_{ABA'B'}=\Phi\o_A\o_B\ob_{A'}\ob_{B'}
$$
where $\Phi$ is a real scalar field of weight $\{-2,-2\}$. All the other
components of the curvature vanish. Substituting into the Bianchi
identities in this formalism ((55)-(65) of reference [2]) gives
$$
\K=0  \label{K} \cr
\S=0  \label{S} \cr
\R=0  \label{R} \cr
$$
together with
$$
\nthorn\bPhi_{22}=0 \label{B1} \cr
\nedthp\bPhi_{22}=\bar\T\bPhi_{22} \label{B2} \cr
$$
Equations \ref{K}--\ref{R} are of course equivalent to the GHP equations
$\kappa=\sigma=\rho=0$ so that $\T$ has the form
$$
\T_{AA'}=\tau\o_A\ob_{A'}
$$
and \ref{B1} and \ref{B2} then give the GHP equations
$$\eqalign{
\thorn\Phi&=0 \cr
\bar{\edth}\Phi&=\bar\tau\Phi \cr
}
$$
Most of the Ricci equations are identically satisfied, with the remaining
equations being
$${\eqalign{
\nthorn\T&=0 \cr
\nedth\T&=\T^2 \cr
\nedthp\T&=\T\bar\T \cr}}
\quad\Leftrightarrow\quad
{\eqalign{
\thorn\tau&=0 \cr
\edth\tau&=\tau^2 \cr
\edthp\tau&=\tau\bar\tau \cr}
}\label{T}
$$
Finally the commutators applied to a general symmetric spinor $\boeta$
reduce to
$$\eqalign{
[\nthornp, \nthorn]\boeta &=-\bar\T\nedth\boeta-\T\nedthp\boeta \cr
[\nthornp, \nedthp]\boeta &=-\bar\T\nthornp\boeta
-\bPhi_{22}(\boeta\cdot\ob)\cr
[\nthornp, \nedth]\boeta &=-\T\nthornp\boeta-\bPhi_{22}(\boeta\cdot\o) \cr
[\nthorn, \nedthp]\boeta&=0 \cr
[\nthorn, \nedth]\boeta&=0 \cr
[\nedthp, \nedth]\boeta&=0 \cr
}\label{com}
$$
Note that the terms involving a contraction with $\o$ are absent
when such a contraction is impossible (as is the case when $\eta$ is a
scalar).
\medskip\noindent

In this particular case it is obvious that the Bianchi and Ricci equations
supply comparatively little information.  Therefore most of the information is
contained in the commutators, and so it is very important that we apply the
commutators in a systematic manner as outlined above, in order to ensure
that all
possible information is obtained. The Bianchi and Ricci equations only
supply us
with  the respective partial tables for
$\tau$ and
$\Phi$ given above, so, in
the next section, we will need to complete these tables (for their
$\nthornp
$ derivative) with two unknown  spinors.  The commutators will first be
applied to a complex weighted scalar $\bar P Q$ formed from $\tau$ and
$\Phi$, and next to one of the two unknown spinors which can be
scaled to be of the type and weight of $\I_A$; then we will apply the
commutators to the
$\{0,0\}$ weighted real scalar $A$ (which is essentially $ \tau \bar \tau$).
These operations will generate new  $\{0,0\}$ weighted scalars and their
respective tables; we will then choose three of these
real scalars, ensuring that they are functionally independent of $A$ and of
each other, and apply the commutators to them.  After a little tidying up we
will obtain an essentially involutive set of tables for four functionally
independent
$\{0,0\}$ weighted real scalars, for one weighted complex scalar and for the
spinor $\I_A$.

By considering the four $\{0,0\}$ weighted real scalars as coordinates, we can
write down, directly from their tables,  the   metric.  The table for the
weighted scalar will supply explicitly the `badly weighted' spin coefficients
$\alpha$, $\beta$, $\epsilon$ and $\gamma$, while the table
for
$\I_A$ yields the other `missing' spin coefficients
$\mu$,
$\nu$, $\lambda$ and $\pi$; of course, our interest is usually just in
obtaining the  metric, and so we would not normally bother evaluating these
other spin coefficients explicitly. However although the information
in these last two tables is not used directly in the calculation of
the metric the tables play an essential role in our intermediate
calculations and in generating
some of the coordinates and their tables.

\section{The integration procedure: the generic case}

\subsection{Preliminary rearranging}

As explained in section 1 the Riemann tensor and its first 
derivative supply three
real scalars  which can easily be rearranged to give one real
zero-weighted $(\tau\bar\tau)$ and two real
weighted scalars, $\Phi$ and $\arg(\tau/\bar\tau)$.
However, simply to keep the presentation of subsequent calculations
to a minimum, it will be convenient to rearrange slightly these three
scalars, and use instead the zero-weighted scalar
$$
A={1\over\sqrt{2\tau\bar\tau}}
\label{A}
$$
and the  weighted scalars
$$
P=\sqrt{\tau\over2\bar\tau}
 \qquad\qquad \hbox{with}\qquad  P\bar P ={1\over 2}\label{P}
$$
$$
Q={\sqrt{\Phi}\over\root 4 \of
{2\tau\bar\tau}}
\label{Q}
$$
of respective weights $P\  \{1,-1\}$ and $Q\  \{-1,-1\}$. 
(We are assuming $\tau\ne 0$, and so each of $A,\ P,\ Q$ will always be defined 
and different from zero.)

These particular
choices  enable us to replace
\ref{T} with the very simple equations
$$
\eqalign{\Ph A & =0\cr
 \D A & =-P\cr
\D'A & =-\bar P\cr
}\label{dA}
$$
and
$$
\eqalign{\Ph P & =0 = \Ph Q \cr
 \D P & =0 = \D Q\cr
\D'P & =0 =\D' Q}\label{dP}
$$

\subsection{Applying commutators to one complex weighted scalar and to one
spinor}

For our integration procedure we begin with a table for the weighted scalar
$(\bar P Q)$, whose weight is $\{-2,0\}$,

$$
\eqalign{\Ph(\bar P Q) & =0\cr
 \D(\bar P Q) & =0\cr
\D'(\bar P Q) & =0\cr
\Ph'(\bar P Q) & =-{Q\over A}{\bf I}\cr}\label{dPQ}
$$
where we have completed the table with a spinor ${\bf I}$, which is as yet
undetermined.  (We have introduced  the weighted factor
${-Q\over A}$ in the above definition for ${\bf I}$ simply for convenience in
later calculations.)

It follows from \ref{thornp.ob} and \ref{thornp.o} that
$$
\eqalign{{\bf I} \cdot \ob & =-{ A\over Q} \bigl(\Ph'(\bar P Q)\bigr)\cdot
\ob
  =-{ A\over Q} \D'(\bar P Q)\cr & =0} \label{I.ob}
$$
$$
\eqalign{{\bf I} \cdot  \o & =-{ A\over Q} \bigl(\Ph'(\bar P Q)\bigr)\cdot  \o
  =-{ A\over Q} \Bigl(\D(\bar P Q)+2\tau(\bar P Q)\Bigr)  \cr & = -1 }
\label{I.o}
$$
Hence ${\bf I} $ is a $(1,0)$ type spinor, and from
$$
\bigl(\Ph'(\bar P Q)\bigr)_{ABA'B'}  =-{Q\over A}{\bf I}_{(A}\o_{B)}
\ob_{A'}\ob_{B'}
$$
we conclude that its weight is  $\{-\1,\0\}$.

So now we have to apply the commutators to the table for $ (\bar P Q)$ which
yields a partial table   for the spinor ${\bf I}$; the complete table can be
written
$$
\eqalign{\Ph {\bf I} & =0\cr
\D{\bf I} & =0\cr
\D'{\bf I} & =0\cr
\Ph'{\bf I} & ={\bar PQ^2\over A}{\bf W}\cr}\label{dI}
$$
where we have completed the table with a spinor ${\bf W}$, which is as yet
undetermined.
It follows that
$$
\eqalign{{\bf W} \cdot \ob & ={ A\over \bar P Q^2} \bigl(\Ph'{\bf
I}\bigr)\cdot
\ob
\cr
 & = 0
}\label{W.ob}
$$
$$
\eqalign{{\bf W} \cdot  \o & =-{ A\over \bar P Q^2} \bigl(\Ph'{\bf
I}\bigr)\cdot
 \o \cr
 & = {1\over Q^2 \bar P^2} {\bf I}
}\label{W.o}
$$
Hence
$$
{\bf W }= -{1\over 2 \bar P^2Q^2} {\bf I}^2 + W \label{W}
$$
where ${\bf W} $ is a $(2,0)$ type spinor of weight   $\{\2,\0\}$, and $W$ is a
zero-weighted complex scalar.

We next have to apply the commutators to the table for ${\bf I}$ which
yields a
partial table for  the
spinor
${\bf W}$; under the substitution \ref{W} we obtain a partial table for the
zero-weighted complex scalar $W$,
$$
\eqalign{\Ph { W} & =0\cr
\D{ W} & =-2P\cr
\D'{ W} & =0\cr
}\label{dW}
$$

\subsection{Finding four coordinate candidates, and applying 
commutators to them}

We have obtained complete tables for the weighted scalar $(\bar P Q)$, and for
the spinor ${\bf I}$, and applied the commutators to each; 
so it remains to obtain complete  tables for four
real zero-weighted scalars, and to apply the commutators to all four of these
scalars.  Clearly $A$ and $W$, for which we already have partial tables, are
obvious candidates.

The complete table for $A$ can be written
$$
\eqalign{\Ph A & =0\cr
 \D A & =-P\cr
\D'A & =-\bar P\cr
\Ph'A & ={Q\over A}{\bf N}\cr}\label{DA}
$$
where we have completed the table with a spinor ${\bf N}$, which is as yet
undetermined.
It follows that
$$
\eqalign{{\bf N} \cdot \ob  ={ A\over  Q} \bigl(\Ph' A\bigr)\cdot
\ob
  = { A\over  Q} \D' A
  =-{ A\over  Q}\bar P
}\label{N.ob}
$$

$$\eqalign{{\bf N} \cdot  \o  ={ A\over  Q} \bigl(\Ph' A\bigr)\cdot
 \o
 = { A\over  Q} \D A
  = - { A\over  Q} P
}\label{N.o}
$$
Hence
$$
{\bf N }= {AP\over Q}{\bf I}+{A\bar P\over Q}\bar {\bf I}+N \label{N}
$$
where ${\bf N} $ is a hermitian $(1,1)$ type spinor of weight   $\{\1,\1\}$,
and $N$
is a zero-weighted real scalar.

We now have to apply the commutators to $A$, which yields a partial table for
${\bf N}$; under the substitution \ref{N} we obtain a partial table for the
zero-weighted real scalar $N$,
$$
\eqalign{\Ph { N} & =-{1\over Q}\cr
\D{ N} & ={\bf \bar I}/Q\cr
\D'{ N} & ={\bf I}/Q\cr
}\label{dN}
$$

Next, considering $W$, we can write down its complete table,
$$
\eqalign{\Ph W & =0\cr
 \D W & =-2 P\cr
\D'W & =0\cr
\Ph'W & ={Q\over A}{\bf Z}\cr}\label{DW}
$$
where we have completed the table with a spinor ${\bf Z}$, which is as yet
undetermined.
It follows that
$$
\eqalign{{\bf Z} \cdot \ob  ={ A\over  Q} \bigl(\Ph' W\bigr)\cdot
\ob
  = 0
  }\label{Z.ob}
$$
$$\eqalign{{\bf Z} \cdot  \o  ={ A\over  Q} \bigl(\Ph' W\bigr)\cdot
 \o
 = -{ 2A P\over  Q}
}\label{Z.o}
$$
Hence
$$
{\bf Z }= {2A P\over Q} {\bf I}+Z \label{Z}
$$
where ${\bf Z} $ is a  $(1,0)$ type spinor of weight   $\{\1,\0\}$, and $Z$
is a zero-weighted complex scalar.

We now have to apply the commutators to $W$, which yields a partial table for
${\bf Z}$; under the substitution \ref{Z} we obtain a partial table for the
zero-weighted complex scalar $Z$,
$$
\eqalign{\Ph { Z} & =-{1\over Q}\cr
\D{ Z} & ={\bf \bar I}/Q\cr
\D'{ Z} & ={\bf I}/Q\cr
}\label{dZ}
$$

Having applied our commutators to (the equivalent of) three real zero-weighted
scalars, we need to identify (at least) one more; clearly $N$ --- which is real
and zero-weight --- is the obvious candidate.  Using \ref{dN}  we can
write down a complete table for $N$,
$$
\eqalign{\Ph { N} & =-{1\over Q}\cr
\D{ N} & ={\bf \bar I}/Q\cr
\D'{ N} & ={\bf I}/Q\cr
\Ph' N & = {Q\over A}{\bf L}}\label{DN}
$$
where we have completed the table with a spinor ${\bf L}$, which is as yet
undetermined.
It follows that
$$
\eqalign{{\bf L} \cdot \ob  ={ A\over  Q} \bigl(\Ph' N\bigr)\cdot
\ob
  = { A\over  Q^2}  {\bf I}
  }\label{L.ob}
$$
$$
\eqalign{{\bf L} \cdot  \o  ={ A\over  Q} \bigl(\Ph' N\bigr)\cdot
 \o
  = { A\over  Q^2} \bar {\bf I}
  }\label{L.o}
$$

Hence
$$
{\bf L }= -{A\over Q^2}{\bf I}\bar {\bf I}+L \label{L}
$$
where ${\bf L} $ is a hermitian $(1,1)$ type spinor of weight   $\{\1,\1\}$,
and $L$ is a zero-weighted real scalar.

We now have to apply the commutators to $N$, which yields a partial table for
${\bf L}$; under the substitution \ref{L} we obtain a partial table for the
zero-weighted real scalar $L$,
$$
\eqalign{\Ph { L} & =0\cr
\D{ L} & =P \bar W\cr
\D'{ L} & =\bar P  W\cr
}\label{dL}
$$

\

So we now have applied our commutators to (the equivalent of) four real
zero-weighted scalars, and {\it providing that these scalars are functionally
independent}, they can be adopted as coordinates. (It will be easier to check
for this functional independence from the scalar operator form of these
tables.)
Furthermore, we have now obtained in an explicit form {\it all} the
information about this class of spaces.
\

However, our  tables for the  zero-weighted scalars $A,W,N$ are
not completely involutive with respect to these zero-weighted scalars, since
they
contain also the zero-weighted scalar functions $L, Z$; but we also know
the defining constraints \ref{dL}, \ref{dZ} on these functions. 
In the next subsection we will
rearrange these tables a little in order that the defining constraints on these
extra scalars have a particularly simple form.

\subsection{Simpler form of the six tables in spinor operators}

Before translating  our tables into the scalar operators, it will be
convenient to write the complex scalars $W,Z$ respectively in their real and
imaginary parts, and by a slight rearranging obtain a  simpler
presentation of the tables.

Putting
$$\eqalign{M & ={1\over 2} (W + \bar W) - A \qquad\qquad
B  = {i\over 2} ( W-  \bar W)\cr
F & ={1\over 2} (Z+\bar Z)-N\qquad\qquad E={i\over 2} ( Z -\bar Z)}$$
we replace the table for (complex) $W$ with the two tables,
$$
\eqalign{\Ph { M} & =0\cr
\D{ M} & =0\cr
\D'{ M} & ={0}\cr
\Ph' M & = {QF \over A}}\label{dM}
$$
$$
\eqalign{\Ph { B} & =0\cr
\D{ B} & =-iP\cr
\D'{ B} & =i\bar P\cr
\Ph' B & = i(P{\bf I}-\bar P \bar {\bf I})+{QE \over A}}\label{dB}
$$
The scalars $E,\  F$  satisfy the simple conditions,
$$
\eqalign{\Ph { E} &    =\Ph F =0\cr
\D{ E} &  = \D F =0\cr
\D'{ E} &  = \D' F= 0\cr
}\label{dE}
$$

Under the substitution
$$
\eqalign{L=S -MA - {1\over 2} (A^2+B^2)
}\label{L}
$$
the table \ref{DN} for $N$ becomes
$$
\eqalign{\Ph { N} & =-{1\over Q}\cr
\D{ N} & ={\bf \bar I}/Q\cr
\D'{ N} & ={\bf I}/Q\cr
\Ph' N & = -{1\over Q}{\bf I}\bar {\bf I}+{Q\over A}\Big(S -MA - {1\over 2}
(A^2+B^2) \Bigl)}\label{dN2}
$$
while $S$ satisfies the simple conditions
$$
\eqalign{\Ph { S} &   =0\cr
\D{ S} &   =0\cr
\D'{ S} &    = 0\cr
}\label{dS}
$$
Finally, alongside the  tables for the
other three coordinate candidates, we
include the table \ref{DA} for $A$ ,
$$
\eqalign{\Ph A & =0\cr
 \D A & =-P\cr
\D'A & =-\bar P\cr
\Ph'A & =P{\bf I} + \bar P \bar {\bf I}+{Q\over A}{ N}\cr}\label{dA2}
$$

For completeness we add the other two tables, although we will not need to use
them in obtaining the metric,
$$
\eqalign{\Ph(\bar P Q) & =0\cr
 \D(\bar P Q) & =0\cr
\D'(\bar P Q) & =0\cr
\Ph'(\bar P Q) & =-{Q\over A} {\bf I}\cr}\label{DPQ}
$$
$$
\eqalign{\Ph {\bf I} & =0\cr
\D{\bf I} & =0\cr
\D'{\bf I} & =0\cr
\Ph'{\bf I} & =-{P\over A } {\bf I}^2 +{\bar PQ^2\over A}( A+M-iB)\cr}
\label{DI}
$$

\subsection{The  tables in terms of scalar operators}

If we identify the spinor ${\bf I}_A$ with the second dyad spinor $\i_A$,
then the four tables for the zero weighted coordinate candidates
$M,N,A,B$ can be easily
translated into the ordinary Newman-Penrose scalar operators,
$$
\eqalign{D { M} & =0\cr
\delta { M} & =0\cr
{\bar \delta}{ M} & ={0}\cr
\Delta M & = {QF \over A}}\label{sdM}
$$
$$
\eqalign{D { N} & =-{1\over Q}\cr
\delta { N} & =0\cr
{\bar \delta}{ N} & =0\cr
\Delta N & = {Q\over A}\Bigl(S -MA - {1\over 2}
(A^2+B^2) \Bigr)}\label{sdN}
$$
$$
\eqalign{D A & =0\cr
 \delta A & =-P\cr
{\bar \delta}A & =-\bar P\cr
\Delta A & ={QN\over A}\cr}\label{sdA}
$$
$$
\eqalign{D { B} & =0\cr
\delta{ B} & =-iP\cr
{\bar \delta}{ B} & =i\bar P\cr
\Delta  B & = {QE \over A}}\label{sdB}
$$

We note again that the four tables for the real zero-weighted scalars 
are not strictly involutive in these scalars; there occur also the 
three real scalars $E,F,S,$  which satisfy
$$
\eqalign{D { E} &  = D F =D S =0\cr
\delta E &  =\delta F= \delta S =0\cr
{\bar \delta}E &  = {\bar \delta} F= {\bar \delta} S =0\cr
}\label{dEFS}
$$
The rearranging which we have just carried out was in order to obtain this
simple version of these conditions. Clearly $\nabla E$, $\nabla F$,
$\nabla S$, are all parallel and each is also parallel to $\nabla M$; hence the
zero-weighted scalars $E,F,S$ can each be assumed to be  an arbitrary
function of
the coordinate candidate $M$ alone (and independent of the coordinate
candidates $A,B,N$).

\subsection{Using coordinate candidates as coordinates}

If we now make the obvious choice of the coordinate candidates as
coordinates
$$
m=M,\quad n=N, \quad a=A, \quad b=B
\label{coord}$$
the above four tables for  the zero-weighted scalars enable us to
write down the tetrad vectors in the coordinates $m,n,a,b$,
$$\eqalign{l^i & = (0,{-1\over Q},0,0)\cr
n^i & = {Q\over a}\Bigl(F,(S-ma-{1\over 2}a^2-{1\over
2}b^2),n,E\Bigr)\cr
m^i & = P(0,0,-1,-i)\cr
\bar m^i & = \bar P(0,0,- 1,i)\cr
}\label{frame}$$
where $E,F,S$ are arbitrary functions of the coordinate $m$. The metric follows
immediately from the equation
$$
g^{ij}=2l^{(i}n^{j)}-2m^{(i}{\bar m}^{j)} \label{metric}
$$
and we see that it does not depend upon the gauge fields $P$ and $Q$.

However we noted in the last subsection that this whole procedure 
is dependent on the condition that the zero-weighted scalars which we 
choose as coordinate candidates are functionally independent. 
However, if we make the assumption that
none of these scalars are constants, then a check of the
determinant formed from the four tables for 
$M,\ N,\ A,\ B$ shows that all four scalars are functionally
independent.  Now it is
easy to check that none of $N,\ A,\ B$ can be constant, but $M$ might be;
therefore the tetrad obtained above is not the most general  that can be
obtained for this class of spacetime. In the next section we will
consider the case where $M$ is constant and we need to find a fourth
coordinate. 

\section{The integration procedure: the complete case}

\subsection{Preliminaries}

In the previous section  we assumed that $M$ was not a constant, so that we
were able to choose it as our fourth coordinate candidate. Next, we should
look at the excluded case where $M$ is a constant. In such a situation, clearly
$F$ is zero, but we still have the possibility of choosing $E$ or $S$ as our
fourth coordinate. Once we make such a choice then we could continue in a
similar
manner as in the last section, building our tables, and hence the tetrad,
around
our four coordinate candidates. However, if {\it all} of the functions $M,E,S$
are constants, then it will {\it not} be possible to find the fourth coordinate
candidate  directly; we emphasise that in such circumstances no additional
independent quantities can be generated by any direct manipulations of the
tables and the commutators. In such a situation we still need a fourth
coordinate
candidate in order to extract the remaining information  from the commutators.
 We shall now
show that by an indirect approach  a fourth coordinate candidate can in fact be
found, so that we can obtain the complete metric as one expression.

\subsection{Finding a fourth coordinate candidate indirectly, 
and extracting all the information from the complete system.}

The results in section 3 up to the end of subsection 3.4 apply; however, when
we are considering tables explicitly for our coordinate candidates we consider
only the  three coordinate candidates $N,\ B,\ A$ while the
zero-weighted scalar
$M$ is not now included as a coordinate candidate.

So, clearly we do not have our full quota of {\it four} coordinate candidates,
but we do not wish to use any of the remaining quantities from the  tables,
since it would involve the additional assumption of that quantity being
non-constant.  However, we know that we have not yet extracted all the
information from the commutators \ref{com}, since they have only been 
applied to {\it three} zero-weighted coordinate  candidates. So
we closely examine the structure of the commutators \ref{com}  
to determine whether they suggest the existence of a
 fourth zero-weighted scalar, functionally independent of the first three
coordinate candidates, whose table is consistent with the commutators.
In fact, we get a strong hint from the previous section, and consider the
possibility of the existence of a real zero-weighted scalar $T $, which
satisfies the table
$$\eqalign{\Ph T & =  0\cr
 \D T & =  0\cr
\D' T & =  0\cr
\Ph' T & =  Q/A
}\label{dT}$$
It is
straightforward to confirm that such a choice is consistent with the
commutators \ref{com} and creates no inconsistency with the other  tables.

\subsection{The scalar tables.}

As in Section 2.5 we identify the spinor ${\bf I}_A$ with the second dyad
spinor
$\i_A$, so that the four tables can be  translated into the ordinary
Newman-Penrose scalar operators.  Therefore, this is simply equivalent
to  replacing the table for $M$ \ref{dM}, with the  table \ref{dT} for $T$,
$$\eqalign{D T & =  0\cr
 \delta T & =  0\cr
{\bar \delta T} & =  0\cr
\Delta T & =  Q/A
}\label{sdT}
$$
while the real zero-weighted quantities
$E,\ M,\ S$ satisfy
$$
\eqalign{D { E} &  = D M =D S =0\cr
\delta E &  =\delta M= \delta S =0\cr
{\bar \delta}E &  = {\bar \delta} M= {\bar \delta} S =0\cr
}\label{sdEMS}
$$
and so $E,\ M,\ S$ are now arbitrary functions of $T$ only. A check on the
determinant formed from the four scalar tables \ref{sdA}, \ref{sdB},
\ref{sdN}, \ref{sdT}, for $A,\ B,\ N,\ T$ respectively, shows that 
all four scalars are functionally independent.

\subsection{Using coordinate candidates as coordinates}

We now make the obvious choice of the coordinate candidates as the coordinates,
$$ t=T,\qquad  n=N, \qquad a=A, \qquad b=B
\label{coord2}$$
where the only coordinate freedom is for $t$ up to an additive constant.
We can write down the tetrad vectors immediately in the $t,n,a,b$
coordinates from the respective tables as
$$\eqalign{l^i & = (0,{-1\over Q},0,0)\cr
n^i & = {Q\over a}\Bigl(1,(S-Ma-{1\over 2}a^2-{1\over
2}b^2),n,E \Bigr)\cr
m^i & = P(0,0,-1,-i)\cr
\bar m^i & = \bar P(0,0,- 1,i)\cr
}\label{frame2}$$
and therefore using \ref{metric} the metric  is given by,
$$g^{ij}=\pmatrix{0 &{-1/ a} & 0&0\cr
{-1/ a} & (-2S+2Ma+a^2+b^2)/a & -{n/ a} & -{E/ a}\cr
0&-{n/ a} & -1&0\cr
0&-{E / a}&0&-1}
\label{metric2}$$
where $E,\ M,\ S$ are arbitrary functions of the coordinate $t$. 
This form now includes the possibility of any of $M$ or $E$ or $S$ 
being constant and represents the most general conformally flat pure
radiation metric. 

In fact this metric is slightly different from that
obtained by Edgar and Ludwig using the original GHP operators [9].
However if we choose to work with a slightly different spin frame
$\{\o_A,\  {\tilde \I}_A\}$ and choice of coordinates $A, \ B,\ T,\
{\tilde N}$, where
$$
{\tilde \I}_A= \I_A + { \bar P Q \omega} o_A
$$
$$
{\tilde N}=N+\omega 
$$
then we may show that the two forms are equivalent by choosing $\omega$
to be a real function of $t$ only which satisfies the condition
$$
\dot \omega +\omega^2+M=0
$$

\section{Karlhede Classification}

In the Karlhede classification one starts by putting the curvature
into a canonical form. For the conformally flat pure radiation
metrics one may choose $\o_A$ so that 
$$
\bPhi_{ABA'B'}=\o_A\o_B\ob_{A'}\ob_{B'} \label{Phi2}
$$
and hence $\Phi=1$. This fixes the boost freedom of the spin frame but
one still has the rotation and null rotation freedom. Because of the
form of the curvature \ref{Phi2}, and the Bianchi identities, the only
terms one obtains from the derivative of the curvature are $\tau$ and
$(\gamma+\bar\gamma)$. Since $\rho=\sigma=\kappa=0$, then $\tau$ is
invariant under null rotations. However it has non-trivial spin weight
and under a rotation
$$
\o_A \gt \e^{\ei \theta/2}\o_A
$$
$\tau$ transforms as
$$
\tau \gt \e^{\ei \theta}\tau
$$
So provided that $\tau \neq 0$ we may fix the rotation freedom by
demanding the $\tau$ is real. Under a null rotation
$$
\i_A \gt \i_A +\bar a \o_A
$$
$(\gamma+\bar\gamma)$ transforms as
$$
(\gamma+\bar\gamma) \gt (\gamma+\bar\gamma) +
a(\alpha+\bar\beta+\bar\tau)  + \bar a(\bar\alpha+\beta+\tau)
$$
So that provided $(\alpha+\bar\beta+\bar\tau) \neq 0$ one can
transform $(\gamma+\bar\gamma)$ to zero. However this does not
completely fix the null rotation freedom because a one-parameter
subgroup of the (two dimensional) group of null rotations maintains
the condition $(\gamma+\bar\gamma)=0$. To fix the frame completely one
has to go to second order or higher. 

In fact the calculations given in this paper show that there is a
canonical choice of frame at second order given by choosing the spin
and boost freedom so that $P=1/\sqrt2$ and $Q=\sqrt A$, and the null
rotation freedom so that $\nthorn'({\bar P}Q)$ is in canonical
form. Note this involves fixing the frame directly at second order and
is slightly different from the frame obtained by first reducing the
null rotations to a one parameter subgroup at first order and fixing
the remaining freedom at second order. In the latter case one has
$$
\i_A=\I_A-{N \over {3\sqrt2}}\o_A
$$
 
Having fixed the frame completely one can calculate scalar invariants
by looking at the components of the covariant derivatives of the
curvature in the canonical frame. We can extract at most four
functionally independent quantities in this way. In the present
approach we do not fix the frame, but apply the invariant differential
operators to the invariant scalar field $A$. We then extract zero
weighted scalar fields using the gauge dependent quantities $P$, $Q$
and $\I_A$. However these are {\it automatically } independent of the
frame and are therefore scalar invariants. Further invariants may be
obtained by applying the invariant operators to these quantities and
extracting zero weighted scalars. In [16], [17] it was shown that all
the functional information obtained in the Karlhede classification  of
type N vacuum spacetimes could be obtained by applying the invariant
operators to $\Psi_4$. In a similar way, for the conformally flat
radiation metrics, all the information may be obtained by applying the
differential operators to $\Phi$. We give explicit expressions for the
terms obtained below.

At zeroth order
%
%
$$
\Phi ={Q^2\over A}
\label{K5}
$$

At first order
$$
\eqalign{\Ph \Phi & =0\cr
\D \Phi & = {P Q^2\over A^2}\cr
\D' \Phi & = {\bar P Q^2\over A^2}
\cr
\Ph' \Phi & = -{ Q^2\over A^2}(3P{\bf I}+3\bar P \bar {\bf I} + {QN\over
A})}
\label{K6}
$$
Using the fact that $P\bar P=1/2$ we may solve these for $A$, $P$ and
$Q$, but $I$ and $N$ are not uniquely determined as one still has left
the gauge freedom of a one parameter subgroup of null rotations.

At second order one obtains
$$
\eqalign{\Ph \D \Phi & =0\cr
\D \D \Phi & = {2 P^2Q^2\over A^2}
\cr
\D'  \D \Phi & = { Q^2\over A^3}\cr
\Ph'\D\Phi & =  -{ PQ^2\over A^3}(3P{\bf I}+5\bar P \bar {\bf I}+{2QN\over A})
\cr
\Ph \Ph'  \Phi & =0\cr
\D \Ph' \Phi & = -{ PQ^2\over A^3}(6P{\bf I}+8\bar P \bar {\bf I}+{3QN\over A})
\cr
\Ph'\Ph' \Phi & = {3 Q^2\over A^3}\Bigl(4P^2{\bf I}^2+5P\bar P {\bf I}\bar {\bf
I} +\bar{P}^2\bar{\bf I}^2\Bigr) +12{ Q^3N\over A^4}(P{\bf I}+\bar P \bar
{\bf I})\cr 
& \qquad\qquad\qquad + { Q^4\over A^4}\Bigl(S-2AM -{5\over2}A^2
+{1\over2}B^2+3N^2/A\Bigr)\cr
}
\label{K8}
$$
We can now solve these equations for $A,\  N,\  P,\  Q$ and ${\I}$ as
well as the scalar combination $S-2AM +{1\over2}B^2$. Thus at second
order we have fixed the frame completely and have three invariant
scalar quantities.

At third order we have the equation
$$
\eqalign{\D (S-2AM
+{1\over2}B^2) & = P(2M-iB)}
\label{K9}
$$
so that we can now solve for $M$, $B$ and $S$. In section 3.6 we
showed that provided $M$ is not constant, then $A$, $B$, $N$ and $M$
provide four functionally independent pieces of information, so that
generically all the information is obtained at third order. However in
the worst case it is possible for all the new terms obtained at third
order (namely $M$, $E$ and $S$) to be constant. In this case it is
necessary to go to 4th order to show that no further relations exist.
That such cases can arise in practice was first shown by Koutras [18]
who showed that a solution of Wils [19] required the fourth covariant
derivative for its invariant classification. Subsequently further
examples which required the fourth derivative were found by Edgar and
Ludwig [9]. Finally it was shown by Skea [19] (and confirmed by the
above calculation) that for all conformally flat pure radiation
solutions, all the information about the spacetime is contained in the
Riemann tensor and its covariant derivatives to no higher than fourth
order. 

\ack 
SBE wishes to acknowledge the financial support of the Swedish Natural
Science Research Council (NFR).

\references

\rf{[1]} \refjl{Machado Ramos M P and Vickers J A 1996}{\PRS}{A 450}{1--17}

\rf{[2]} \refjl{Machado Ramos M P and Vickers J A 1996}{\CQG}{13}{1579--87}

\rf{[3]} \refjl{Geroch R, Held A and Penrose R 1973}{\JMP}{14}{874--81}

\rf{[4]} \refjl{Held A 1974}{\CMP}{37}{311--26}

\rf{[5]} \refjl{Held A 1975}{\CMP}{44}{211--22}

\rf{[6]} \refjl{Held A 1974}{\GRG}{7}{177--98}

\rf{[7]} \refjl{Edgar S B 1992}{\GRG}{24}{1267--95}

\rf{[8]} \refjl{Edgar S B and Ludwig G 1997}{\GRG}{29}{19--59}

\rf{[9]} \refjl{Edgar S B and Ludwig G 1997}{\GRG}{29}{1309--28}

\rf{[10]} \refjl{Karlhede A and Lindstr\"om U 1983}{\GRG}{15}{597--610}

\rf{[11]} \refjl{Bradley M and Karlhede A 1990}{\CQG}{7}{449--463}

\rf{[12]} \refjl{Bradley M and Marklund M 1996}{\CQG}{13}{3021--37}

\rf{[13]} \refjl{Marklund M 1997}{\CQG}{14}{1267--84}

\rf{[14]} {Marklund M and Bradley M 1998}{ \it Preprint}

\rf{[15]} {Machado Ramos M P 1997}{\it PhD Thesis, University of Southampton}

\rf{[16]} \refjl{Machado Ramos M P and Vickers J A 1996}{\CQG}{13}{1589--99}

\rf{[17]} \refjl{Machado Ramos M P 1998}{\CQG}{15}{435-43}

\rf{[18]} \refjl{Koutras A 1992}{\CQG}{9} {L143-5}

\rf{[19]} \refjl{Skea J E F 1997} {\CQG}{14}{2393-404}

\bye